\documentclass[prl,twocolumn,floatfix,superscriptaddress]{revtex4-1}
\usepackage{dcolumn,amsmath}
\usepackage{graphicx}
\usepackage{bm}
\usepackage{hyperref}
\usepackage{xcolor}

\usepackage[utf8]{inputenc}
\usepackage[T1]{fontenc}

\usepackage{tabularx}
\usepackage{braket}

\begin{document}

\title{Global and local approaches to population analysis: \\ bonding patterns in superheavy element compounds}

\author{Alexander V. Oleynichenko}
\email{alexvoleynichenko@gmail.com}
\affiliation{Petersburg Nuclear Physics Institute named by B.P.\ Konstantinov of National Research Center ``Kurchatov Institute'' (NRC ``Kurchatov Institute'' - PNPI), 1 Orlova roscha, Gatchina, 188300 Leningrad region, Russia}
\affiliation{Department of Chemistry, M.V. Lomonosov Moscow State University, Leninskie gory 1/3, Moscow, 119991~Russia}
\homepage{http://www.qchem.pnpi.spb.ru}

\author{Andr\'ei V. Zaitsevskii}
\affiliation{Petersburg Nuclear Physics Institute named by B.P.\ Konstantinov of National Research Center ``Kurchatov Institute'' (NRC ``Kurchatov Institute'' - PNPI), 1 Orlova roscha, Gatchina, 188300 Leningrad region, Russia}
\affiliation{Department of Chemistry, M.V. Lomonosov Moscow State University, Leninskie gory 1/3, Moscow, 119991~Russia}

\author{Stepan~Romanov}
\affiliation{Skolkovo Institute of Science and Technology, Skolkovo Innovation Center, Building 3, Nobelya st., Moscow, 143026 Russia}

\author{Leonid V.\ Skripnikov}
\email{leonidos239@gmail.com}
\affiliation{Petersburg Nuclear Physics Institute named by B.P.\ Konstantinov of National Research Center ``Kurchatov Institute'' (NRC ``Kurchatov Institute'' - PNPI), 1 Orlova roscha, Gatchina, 188300 Leningrad region, Russia}
\affiliation{Saint Petersburg State University, 7/9 Universitetskaya nab., St. Petersburg, 199034 Russia}

\author{Anatoly V. Titov}
\affiliation{Petersburg Nuclear Physics Institute named by B.P.\ Konstantinov of National Research Center ``Kurchatov Institute'' (NRC ``Kurchatov Institute'' - PNPI), 1 Orlova roscha, Gatchina, 188300 Leningrad region, Russia}
\affiliation{Saint Petersburg State University, 7/9 Universitetskaya nab., St. Petersburg, 199034 Russia}

\begin{abstract} 
Relativistic effective atomic configurations of superheavy elements Cn, Nh and Fl and their lighter homologues (Hg, Tl and Pb) in their simple compounds with fluorine and oxygen are determined using the analysis of local properties
of molecular Kohn-Sham density matrices in the vicinity of heavy nuclei.
The difference in populations of atomic spinors with the same orbital angular momentum and different total angular momenta is demonstrated to be essential for understanding the peculiarities of chemical bonding in superheavy element compounds.
The results are fully compatible with those obtained by the relativistic iterative version of conventional
projection analysis of global density matrices. 
\end{abstract}

\maketitle

\section{Introduction}

In the last two decades transactinide elements with atomic numbers exceeding 110 had been synthesized \cite{Oganessian2009}. In spite of several successful chemical experiments with some of these species \cite{Guseva2005,Eichler2013,Superheavy}, the bulk of information on their chemical properties is obtained via electronic structure modelling of their compounds. Due to very strong relativistic effects the bonding pattern in these compounds can be rather exotic. Unfortunately, modern electronic structure theory provides very few reliable tools for qualitative interpretation of relativistic models in ``chemical'' terms.
A promising way to describe chemical bonds is to define the so-called effective state 
(effective electronic configuration) of atom in compound. The intuitive definition 
of the effective electronic configuration $\{{\bf n}^{\rm A}\}$ of a particular atom A operates 
with the fractional occupancies $\{n_{lj}\}$ of its valence subshells with definite spatial ($l$) and total ($j$) angular momenta:

\begin{equation}
\label{eq:def_eff_conf}
\{{\bf n}^{\rm A}\} = \{ n_{s_{1/2}}^{\rm A}, n_{p_{1/2}}^{\rm A}, n_{p_{3/2}}^{\rm A}, n_{d_{3/2}}^{\rm A}, n_{d_{5/2}}^{\rm A}, ... \}
\end{equation}

The most popular approaches to determine the sets of occupancies are based on the analysis of the whole (global) one-electron density matrix using some atom-centered functions to split this entity into atomic contributions.
The simplest technique of this type is a straightforward relativistic generalization of Mulliken population analysis 
\cite{Mulliken1955, Pershina2002}.
However, results obtained using this approach are not highly reliable because of their critical dependence on the basis set 
used to discretize the electronic Hamiltonian \cite{Jensen}.

The problem is partially solved by introducing auxiliary restricted (``minimal'') bases of some pre-constructed reference atomic spinors and using these bases to approximately re-expand one-electron density matrices.
One of population analysis techniques of this type is the relativistic implementation of projection analysis (PA)  \cite{Dubillard2006}.
The reference spinors are normally generated as solutions of some SCF-like problem for the free atom and therefore depend on the assumed reference atomic configurations. This gives rise to a certain arbitrariness of the computational scheme and the resulting sets of occupation numbers $\{n^{\rm A}_{lj}\}$.

An alternative approach to determine effective configurations, employing recently proposed Atoms-in-Compounds (AiC) theory \cite{TitovLomachuk2014,Zaitsevskii2016}, consists in the local analysis of molecular density matrices in the vicinities of heavy nuclei and simulating their basic features in the fractional-occupancy calculations of the corresponding free heavy atoms. In contrast to the case of global analysis, no need appears to split the density matrix into atomic contributions. Furthermore, resulting configurations are directly related to certain measurable (spectral) characteristics of compounds, namely, to those associated with effects and processes localized in core domains of heavy atoms, such as  positions of X-ray emission lines.

The first and so far the only systematic application of the Atoms-in-Compound technique for the determination of valence states of heavy atoms has been recently reported for actinides (Pu -- Cf) in their higher oxide molecules \cite{Zaitsevskii2016}. This study was restricted to the scalar relativistic approximation which performs satisfactory for this class of compounds due to moderate spin-orbit splittings of $f$-subshells. In contrast, spin-dependent relativistic effects in the compounds of elements 112 -- 114 are huge, thus requiring to treat separately the subshells with different couplings of spatial and spin angular momenta.
Strong relativistic contraction and stabilization of $s$ and $p_{1/2}$ subshells and secondary destabilization of the shells with higher angular momentum lead to dramatic dissimilarities in the chemical behavior of these elements and their lighter homologues. The chemistry of copernicium (E112) and flerovium (E114) is of particular interest, since their atoms have quasi-closed-shell ground-state electron configurations, $6d^{10} 7s^2$ for Cn and $6d^{10}7s^2 7p_{1/2}^2$ for Fl. This fact should lead to a relative inertness of these elements in most chemical interactions, which has been confirmed both theoretically \cite{Pershina2009} and experimentally by thermochromatography on gold surfaces \cite{Superheavy}. Similar experiments on nihonium (Nh, E113) have shown that its chemical compounds can be volatile; the species observed in experiments were attributed to either atomic nihonium or its hydroxide NhOH \cite{Dmitriev2014}. The next step towards understanding the unwonted chemical bonding in molecules of Cn and Fl was made recently with the projection analysis technique \cite{Oleynichenko2017}.

The present paper reports the study of qualitative and quantitative differences in electronic structures of simple oxygen- and fluorine-containing compounds of superheavy elements and their lighter homologues,
Cn \emph{vs} Hg, Tl \emph{vs} Nh and Fl \emph{vs} Pb, through evaluating relativistic effective configurations of heavy atoms in compounds by both Atom-in-Compound (AiC) and, if necessary, projection population analysis techniques.
The next section summarizes the main features of the AiC-based procedure of subshell population calculations,
focusing on its advantages and shortcomings with respect to the global analysis of density matrices. 
Finally, the results of application of both (global and local) approaches to some molecules containing heavy and superheavy atoms are presented and peculiarities in the chemical bonding in these molecules are discussed.

\section{Outline of theory}

Let us start with recalling the main features of the relativistic projection analysis (PA) technique \cite{Dubillard2006} which seems to be one of the most reliable implementation of the global approach. This technique implies the approximate re-expansion of density matrices in the minimal or nearly minimal set of atomic spinors  obtained as solutions of some SCF-like problem for free atoms. 
Technically, this operation is performed \emph{via} projecting the reference spinors onto the occupied (or natural) molecular spinors. Fractional occupancies of these new basis spinors are calculated similarly to Mulliken analysis and then summed over atomic subshells to get the effective configuration (\ref{eq:def_eff_conf}).
  
Due to the use of very restricted reference spinor sets, a certain fraction of electron density is not assigned to any atom. The problem can be solved by the passage from ``genuine'' atomic spinors to so-called "intrinsic" atomic spinors \cite{Knizia2013}. However, this distorts clear physical meaning of effective atomic configuration, since the "intrinsic" atomic spinors do not possess any definite angular momentum with respect to the nucleus of the corresponding atom.

As has been already mentioned, the resulting fractional occupancies ${n_{lj}^{\rm A}}$ depend on the particular choice of reference atomic spinors, which are in turn defined by the SCF-like problems used to generate these spinors (normally by the configurations of reference atoms) chosen with a certain degree of arbitrariness.
In order to avoid this arbitrariness, in the previous paper \cite{Oleynichenko2017} we proposed a slightly modified version of the PA technique, based on the use of fractional-occupancy reference atomic configurations coinciding with the effective configurations of atom in compound; these configurations are determined by iterations until self-consistency. Thus the iterative PA still uses the configuration-dependent reference atomic functions, only avoiding the arbitrariness of the choice of reference configuration.
The proposed method is suitable at least for atoms bearing positive net charges. Otherwise, atomic calculations for negative-charged ion can lead to unphysical spatially blurred atomic spinors.

An alternative strategy we use in the present work consists in the local analysis of molecular density matrix in the vicinity of chosen heavy nucleus.
The effective configuration of a heavy atom A in a molecule
is defined as a set of fractional occupation numbers in SCF (HF or Kohn--Sham) calculations of the free or
spherically constrained atom A which yields atomic density matrix near
the nucleus fitting its molecular counterpart in some averaged manner. 

To specify the fitting criteria, consider the partial-wave expansion of valence and subvalence shell contribution to the molecular density matrix $\rho_v({\bf{r}}|{\bf{r}}')$.
In the vicinity of the chosen nucleus a remarkable accuracy can be attained 
with only one radial function $f_{lj}(|{\bf r}|)$ per partial wave \cite{TitovLomachuk2014}:

\begin{eqnarray}
\rho^v({\bf r}|{\bf r}')\approx\sum_{ljm,l'j'm'}\Delta_{ljm,l'j'm'}
Y_{ljm}\left(\frac{\bf r}{|{\bf r}|}\right)Y_{l'j'm'}^\dag\left(\frac{{\bf r}'}{|{\bf r}'|}\right) \times \nonumber \\ \times f_{lj}(|{\bf r}|)f_{l'j'}(|{\bf r}'|), |{\bf r}'|,|{\bf r}'|\le R_c \nonumber \\
\end{eqnarray}
\label{pw}

$R_c$ is the so-called proportionality radius \cite{TitovLomachuk2014}.
Here we suppose that the origin is placed at the center of the nucleus A, $Y_{ljm}$ denotes the spherical spinor
with the total angular momentum projection $m$ and $\Delta_{ljm,l'j'm'}$ are numerical coefficients. 
The functions $f_{lj}$  can be obtained as appropriately normalized
radial parts of any high-energy occupied (valence or subvalence)
or low-lying virtual orbitals of any low-energy configuration of the A atom; all these radial parts are mutually proportional at small distances
from the nucleus and thus yield nearly the same $f_{lj}$ provided $R_c$ is small enough ~\cite{Proportion}.
This holds both for all-electron calculations and when so-called
``hard'' pseudopotentials are used. Assuming 
${\bf r}={\bf r}'$ and integrating 
over angular variables
$\Omega = {\bf r} / {\bf |r|}$, one can obtain the expression 
for the spherically averaged (sub)valence electron density $\overline{\varrho}^v(|{\bf r}|)=\int \rho^v({\bf r}|{\bf r}) d\Omega$
split into the partial $lj$-wave contributions,
\begin{equation}
\overline{\varrho}^v(|{\bf r}|)=\sum_{lj}\overline{\varrho}^v_{lj}(|{\bf r}|),\quad \overline{\varrho}^v_{lj}(|{\bf r}|)
= \sum_{m}\Delta_{ljm,ljm}\left|f_{lj}(|{\bf r}|)\right|^2
\end{equation}
If the functions $f_{lj}$ are normalized by the condition $\int_{|{\bf r}| \le R_c}\left|f_{lj}(|{\bf r}|)\right|^2 d{\bf r}=1$, the coefficients $q_{lj}=\sum_{m}\Delta_{ljm,ljm}$ (partial wave charges) are the contributions 
 from the subshells with orbital angular momentum $l$ and total angular momentum $j$
 to the electronic charge within the sphere of radius $R_c$. 
Now we can formulate the fitting criteria as the coincidence of $q_{lj}^{\rm A}$ values for the molecular density
matrix expansion near the nucleus A 
with those obtained for free or spherically constrained  atom obtained in fractional-occupancy SCF calculations:

\begin{equation}
\label{eq:qmol_qat}
q_{lj}^{\rm A}({\rm mol}) = q_{lj}^{\rm A}(n_s^{\rm A}, n_{p_{1/2}}^{\rm A}, n_{p_{3/2}}^{\rm A}, ...)({\rm atom})\quad \forall\ l, j 
\end{equation}
These highly nonlinear equations are solved with respect to the set of $lj$-shells occupation numbers $\{n_{lj}^{\rm A}\}$ of a free atom A.

It should be noticed that the existence of solutions of Eq. (\ref{eq:qmol_qat}) is not guaranteed. One frequently encounters the situations where the partial wave charges for a free atom with any sets of fractional occupancies are  lower than their counterparts for the atom in a compound. This is apparently related to the fact that an atom in compound is in a sense spatially constrained and the constrains should result in pushing the electron density toward the nucleus. Therefore the use of a confined, rather than free, atom in fractional-occupancy calculations seems both physically grounded and advantageous from the computational point of view, enabling one to
get solutions of Eq. (\ref{eq:qmol_qat}) in difficult cases. We placed the atom into the center of potential well created by an uniformly charged sphere with the radius coinciding with the separation between this atom and its nearest neighbour in the compound. The sphere charge was chosen in such a way that the sum of atomic spinor populations satis\-fying Eq. (\ref{eq:qmol_qat}) coincided with the electronic density integrated over the Bader domain of the heavy atom, i.e. AiC analysis provided the same atomic net charges as the Bader scheme. It is worth noting that the latter requirement forces automatically the equality of the sum of all AiC occupancies and the total number of electrons which is not guaranteed in the case of using free single atoms.

Let us list the main advantages of the local (AiC) analysis over the conventional (global) approach:
\begin{itemize}
\item the problem of assigning contributions to the electronic density / density matrix to a certain atom does not appear;
\item no arbitrary atomic reference state or function should be defined;
\item density matrix fitting criteria is formulated in terms of partial $lj$-wave charges, i.e. the entities 
directly related to experimentally observable ``core'' properties (chemical shifts in X-ray emission spectra, hyperfine structure constants and some others) \cite{TitovLomachuk2014}.
\end{itemize}

\section{Computational details}

All calculations were performed in the two-component relativistic DFT framework with the PBE0 exchange-correlation functional \cite{Adamo1999}. It could be seen from the Table 1 that DFT/PBE0 predictions of the energy characteristics of covalent bonds in these molecules agree reasonably with available experimental and theoretical results. The reliability of the PBE0 functional for this class of heavy element compounds was also demonstrated in \cite{Demidov2015}.
We replaced core electrons with spin-orbit semilocal relativistic pseudopotentials (PPs) by Mosyagin et al \cite{Mosyagin2010} (60-e PPs for Hg, Tl, Pb and 92-e PPs for Cn, Nh, Fl). 
Valence and subvalence spinors of heavy atoms were represented using flexible uncontracted basis set, designed specially for DFT calculations with the employed PPs \cite{pnpi_site}. 
For H, O and F atoms we used the Def2-TZVPD basis set \cite{Rappoport2010} with sligtly modified diffuse functions. The quality of the chosen bases is sufficient for neglecting basis set superposition errors at the DFT level, which were estimated as a few thousandth of electronvolt for diatomic species.
Population analysis was performed for equilibrium molecular geometries (see Table 1). 
Geometry optimizations and one-step projection analysis calculations were carried out using the DIRAC15 package \citep{DIRAC15}. An iterative projection analysis procedure \cite{Oleynichenko2017} was implemented on the top of DIRAC.

In order to find trends in variations of typical bond strength in pairs Hg -- Cn and Pb -- Fl we also evaluated dissociation energies of all studied molecules. 
Kramers-unrestricted RDFT calculations were carried out using the code \cite{Wullen2010}.

The Bader population analysis was performed with the program \cite{Henkelman2009,SanvilleHenkelman2007}. The code used to perform density matrix re-expansion in the vicinity of heavy nuclei was developed by L. Skripnikov \cite{Skripnikov2015}. We assumed the ``proportionality radius'' value $R_c = 0.5$ a.u. With this choice, the normalized $f_{lj}$ functions corresponding to both valence and subvalence atomic spinors obtained in calculations for various reference atomic configurations were similar enough to ensure the stability of the resulting AiC atomic spinor populations in molecule up to a few thousandths.

\section{Results and discussion}

In order to study the peculiarities of effective atomic configurations in superheavy element compounds, both iterative projection analysis and Atom-in-Compound technique were applied to molecules of fluorides and oxides $\rm MF_2$, $\rm MF_4$, $\rm MO$ (M = Hg, Pb, Cn, Fl) as well as hydroxides MOH (M = Tl, Nh). Equilibrium geometry parameters and effective configurations of heavy atoms are summarized in Tables 1 and 2.

\begin{table*}[t!]
\caption{\label{tab:general} \scriptsize Equilibrium geometries and dissociation energies $E_{M-X}$ of studied molecules. $E_{M-X}$, $X = \rm O,OH,F$, are the adiabatic energies for $\rm \frac{1}{n} MF_n \rightarrow \frac{1}{n} M + F$, $\rm MOH \rightarrow M + OH$ and $\rm MO \rightarrow M + O$ reactions (estimate without zero potential energy). RPP: relativistic pseudopotential; DKH: Douglas-Kroll-Hess Hamiltonian.
}
\begin{tabularx}{\textwidth}{XXXXXXll}
\hline
\hline
         &          &                &                   & \multicolumn{3}{c}{dissociation energy $E_{M-X}$, eV} \\
Molecule & Symmetry & $R_{M-X}$, \AA & Angle, $^{\circ}$ & \multicolumn{1}{l}{Present work} & \multicolumn{2}{l}{Other sources} & \multicolumn{1}{l}{Ref.} \\
\hline
\\
\multicolumn{8}{c}{\it Group 12 elements} \\
\\
$\rm HgO$    &  $C_{\infty v}$  &  1.88  &    --   &  0.04  &  0.17 & DKH/CCSD(T) & \cite{SheplerPeterson2003}  \\
$\rm CnO$    &  $C_{\infty v}$  &  1.86  &    --   &  0.20  &  --  & & \\
$\rm HgF_2$  &  $D_{\infty h}$  &  1.90  &  180.0  &  2.72  &  2.69 & RPP/CCSD(T) & \cite{Liu1999} \\
$\rm CnF_2$  &  $D_{\infty h}$  &  1.93  &  180.0  &  2.16  &  1.89 & RPP/CCSD(T) & \cite{Seth1997} \\
$\rm HgF_4$  &  $D_{4h}$        &  1.88  &  180.0  &  1.89  &  1.80 & RPP/CCSD(T) & \cite{Liu1999} \\
$\rm CnF_4$  &  $D_{4h}$        &  1.94  &  180.0  &  1.75  &  1.51 & RPP/CCSD(T) & \cite{Seth1997} \\
\\
\multicolumn{8}{c}{\it Group 13 elements} \\                     
\\
$\rm TlOH$   &  $C_{s}$         &  2.12  &  118.8  &  3.16  &  3.42 & Exptl. & \cite{CRC}  \\
$\rm NhOH$   &  $C_{s}$         &  2.23  &  108.9  &  1.66  &  1.95 & RPP/CCSD(T) & \cite{Demidov2015}  \\
\\
\multicolumn{8}{c}{\it Group 14 elements} \\                       
\\
$\rm PbO$    &  $C_{\infty v}$  &  1.90  &    --    &  3.68  &  3.96 & Exptl. & \cite{CRC}  \\
$\rm FlO$    &  $C_{\infty v}$  &  2.04  &    --    &  0.86  &  1.26 & RPP/CCSD(T) & \cite{Liu2001} \\
$\rm PbF_2$  &  $C_{2v}$        &  2.03  &  95.8   &  4.16  &  4.09 & Exptl. & \cite{ChaseTables1998}  \\
$\rm FlF_2$  &  $C_{2v}$        &  2.16  &  97.4   &  2.20  &  2.30 & RPP/CCSD(T) & \cite{Seth1998}  \\
$\rm PbF_4$  &  $T_d$           &  1.97  &  109.5  &  3.21  &  3.43 & Exptl. & \cite{PbF4expt}  \\
$\rm FlF_4$  &  $T_d$           &  2.12  &  109.5  &  1.26  &  1.49 & RPP/CCSD(T) & \cite{Seth1998} \\
\\
\hline
\hline
\end{tabularx}
\end{table*}

One readily notices that iterative PA results semi-quantitatively coincide with the predictions of the AiC analysis.
The largest discrepancies between AiC and iterative PA based occupancies are observed for the valence $p_{3/2}$ atomic spinors of Hg and Cn. These spinors computed for free atom configurations with weakly occupied valence $p$-shell are too diffuse to describe efficiently the deformation of $s$-subshell in molecules, so the results of PA are not expected to be as reliable as in other cases. In the present version of AiC-based analysis, no similar problem is encountered because the fractional-occupancy single atom is spatially constrained.

This allows us to discuss bonding features on the basis of both population analysis techniques without distinguishing between them.

\begin{table*}[t!]
\caption{\label{tab:aic_occ} \scriptsize Effective relativistic configurations of heavy atom M according to the proposed Atom-in-Compound technique (corresponding iterative projection analysis results are given in parentheses). For MO and $\rm MF_2$ (M = Hg, Cn, Pb, Fl) we cite here the projection analysis results from our previous paper \cite{Oleynichenko2017}. }
\begin{tabularx}{\textwidth}{XXXXXX}
\hline
\hline
         & \multicolumn{5}{c}{Subshell occupancies $n_i$}            \\
Molecule & $s_{1/2}$ & $p_{1/2}$ & $p_{3/2}$ & $d_{3/2}$ & $d_{5/2}$ \\
\hline
\\
\multicolumn{6}{c}{\it Group 12 elements} \\
\\
$\rm HgO $	& 1.26 (1.28)  &  0.17 (0.16)  &  0.18 (0.13)  &  3.94 (3.96)  &  5.77 (5.84)  \\
$\rm CnO $	& 1.65 (1.70)  &  0.33 (0.31)  &  0.23 (0.08)  &  3.92 (3.94)  &  5.30 (5.45)  \\
$\rm HgF_2$	& 0.92 (0.91)  &  0.16 (0.10)  &  0.30 (0.13)  &  3.98 (3.92)  &  5.85 (5.80)  \\
$\rm CnF_2$	& 1.35 (1.48)  &  0.25 (0.15)  &  0.35 (0.08)  &  3.93 (3.88)  &  5.38 (5.46)  \\
$\rm HgF_4$	& 0.69 (0.63)  &  0.18 (0.16)  &  0.30 (0.22)  &  3.67 (3.79)  &  5.08 (5.38)  \\
$\rm CnF_4$	& 1.01 (1.08)  &  0.31 (0.27)  &  0.38 (0.17)  &  3.68 (3.78)  &  4.60 (4.98)  \\
\\
\multicolumn{6}{c}{\it Group 13 elements} \\
\\
$\rm TlOH$	& 1.76 (1.91)  &  0.24 (0.24)  &  0.31 (0.25)  &  4.00 (3.99)  &  6.00 (5.99)  \\
$\rm NhOH$	& 1.91 (1.95)  &  0.43 (0.47)  &  0.19 (0.11)  &  3.99 (3.99)  &  5.95 (5.97)  \\
\\
\multicolumn{6}{c}{\it Group 14 elements} \\
\\
$\rm PbO $	& 1.74 (1.87)  &  0.61 (0.67)  &  0.62 (0.61)  &  4.00 (3.99)  &  6.04 (5.99)  \\
$\rm FlO $	& 1.89 (1.92)  &  1.07 (1.16)  &  0.33 (0.26)  &  3.98 (3.98)  &  5.99 (5.97)  \\
$\rm PbF_2$	& 1.70 (1.87)  &  0.31 (0.37)  &  0.40 (0.44)  &  4.05 (4.00)  &  6.05 (5.99)  \\
$\rm FlF_2$	& 1.88 (1.94)  &  0.51 (0.61)  &  0.28 (0.26)  &  4.00 (3.99)  &  5.97 (5.98)  \\
$\rm PbF_4$	& 0.88 (0.92)  &  0.29 (0.36)  &  0.44 (0.49)  &  4.00 (3.99)  &  5.99 (5.98)  \\
$\rm FlF_4$	& 1.38 (1.42)  &  0.45 (0.51)  &  0.41 (0.32)  &  4.00 (3.99)  &  5.95 (5.96)  \\
\\
\hline
\hline
\end{tabularx}
\end{table*}

Let us recall that in the non-relativistic (and scalar-relativistic) models the formation of one or two covalent bonds by an atom in a closed shell molecule implies the presence of one or two unpaired electrons respectively above the closed shell in the free atom. In case of purely covalent bonding, the overall populations of involved atomic orbitals remain the same upon the bond formation. If the atomic subshell population approaches that of the filled shell, the same holds for hole populations.
Therefore the deviation of the atomic $l$-subshell occupancy from  0 (for low populations) or from $2(2l+1)$ (for high populations) in a closed-shell molecule is related to the number of single covalent bonds formed by the atom. It is natural to suppose that such deviation should correlate with the covalent contribution to chemical bonding  in the case of fractional populations as well.

For heavy element compounds these considerations should be modified.
Let us recall that the atomic two-component spinors are admixtures of spin-up and spin-down components.
Provided that we have a single electron on the $p$-subshell, the formation of a true $\sigma$-bond along the interatomic axis (let us call it $z$) implies the participation of an electron on the $p\sigma{}$ (or, the same, $l=1$, $m_l = 0$) component. However, neither any of the  $p_{1/2}$ spinors  nor any combination of these spinors is dominated by this component. The same holds for the $p_{3/2}$ spinors; to get a pure $p\sigma{}$ function, $p$ spinors with different $j$ are to be combined~\cite{Lee2004}. Such mixing is 
hindered by large differences of energies and spatial distributions of the $j$-subshells with the same $l$,
as occurs in superheavy elements.
A large difference between the populations (per spinor) of the shells with the same $l$ and different $j$ indicates inefficient mixing and therefore 
weak covalent $\sigma$-bonding. A similar argumentation is applicable to higher-$l$ shells.

\begin{figure}
\label{fig:HgCnF2}
\center
\caption{
Effective subshell occupancies per spinor for heavy atoms
in group 12 elements fluorides $\rm MF_2$ and $\rm MF_4$ (M = Hg, Cn), obtained via the AiC analysis. Values are given relative to the filled subvalence shell, negative d-occupancies correspond to subvalence d-shell holes.
}
\includegraphics[scale=.4]{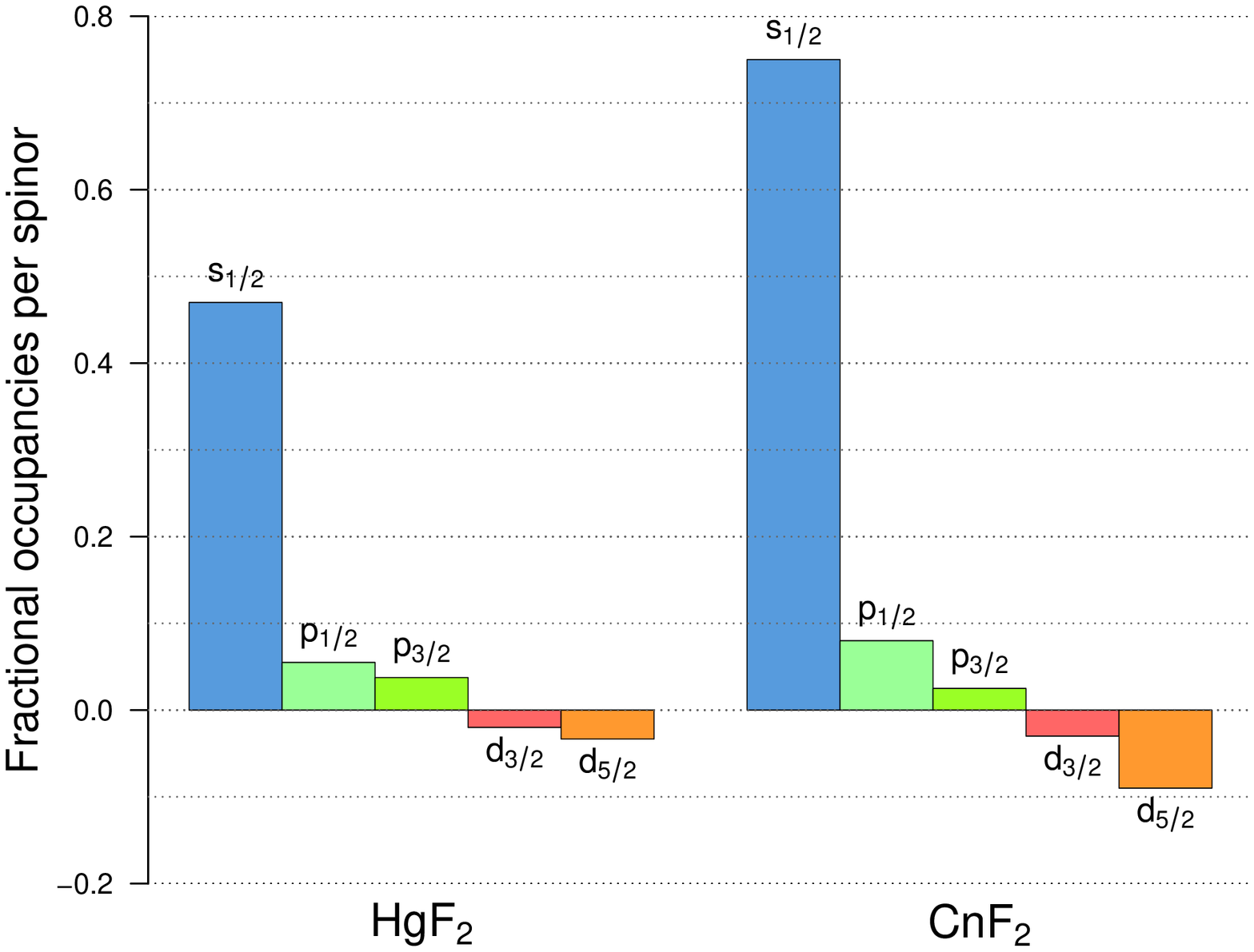}
\includegraphics[scale=.4]{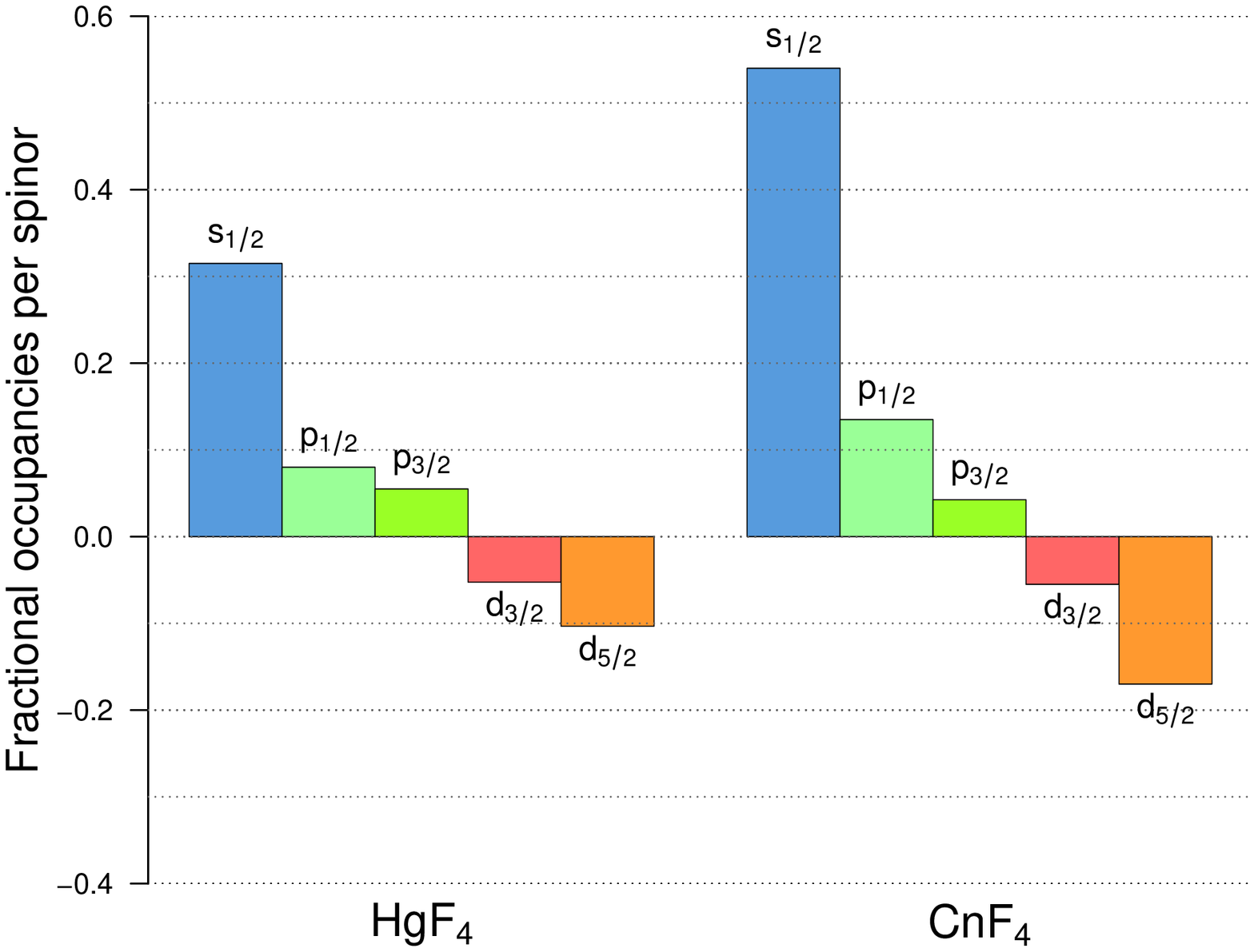}
\end{figure}

In case of compounds of group 12 elements (Hg and Cn) chemical bonds are formed mainly by $s$ electrons (see Figure 1). We may assert that the covalent component of chemical bonds is stronger in $\rm HgF_2$ than in $\rm CnF_2$ since in the former case the fractional occupancy of the $s$-shell approaches 1.
Furthermore, though hole population of the Cn $d_{5/2}$ subshell is significant, it does not strongly contribute to $\sigma$-bonding since the very small hole population of $d_{3/2}$ indicates the inefficiency of $d_{5/2}-d_{3/2}$ mixing required to form $d\sigma$ components. However, $d$-subshells seems not to be as inert as in the case of Hg where they are nearly filled, that characterizes Cn as a real transition element.
It should be underlined that the smallness of HgO and CnO dissociation energies do not indicate the weakness of chemical bonds. This rather exotic situation is related to the fact that the molecular ground states correlate with excited states of separated atoms \cite{SheplerPeterson2003}, so that the Hg--O bond breaking energy within the single-electronic-state model (corresponding to ``spectroscopic'' dissociation energy) is rather large (ca. 2.8 eV, \cite{ChaseTables1998}). This blocks the possibility of using ground-state data providing the ``thermodynamic'' dissociation energy for searching the correlations between the bond strength and effective atomic configuration.

\begin{figure}
\label{fig:TlNhOH}
\center
\caption{Effective subshell occupancies per spinor for heavy atoms
in group 13 elements hydroxides TlOH and NhOH, obtained via the AiC analysis. Values are given relative to the filled subvalence shell.}
\includegraphics[scale=.4]{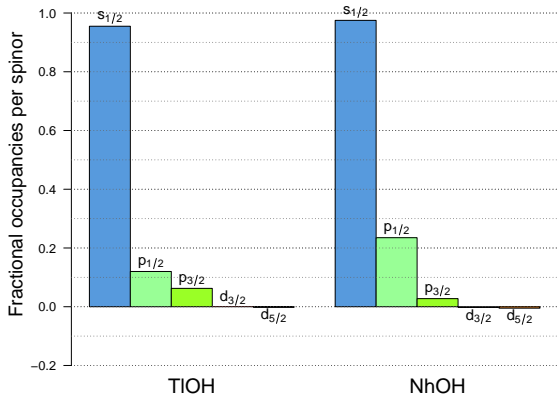}
\end{figure}

\begin{figure}
\label{fig:PbFlF2}
\center
\caption{Effective subshell occupancies per spinor for heavy atoms
in group 14 elements difluorides $\rm PbF_2$ and $\rm FlF_2$, obtained via the AiC analysis. Values are given relative to the filled subvalence shell.}
\includegraphics[scale=.4]{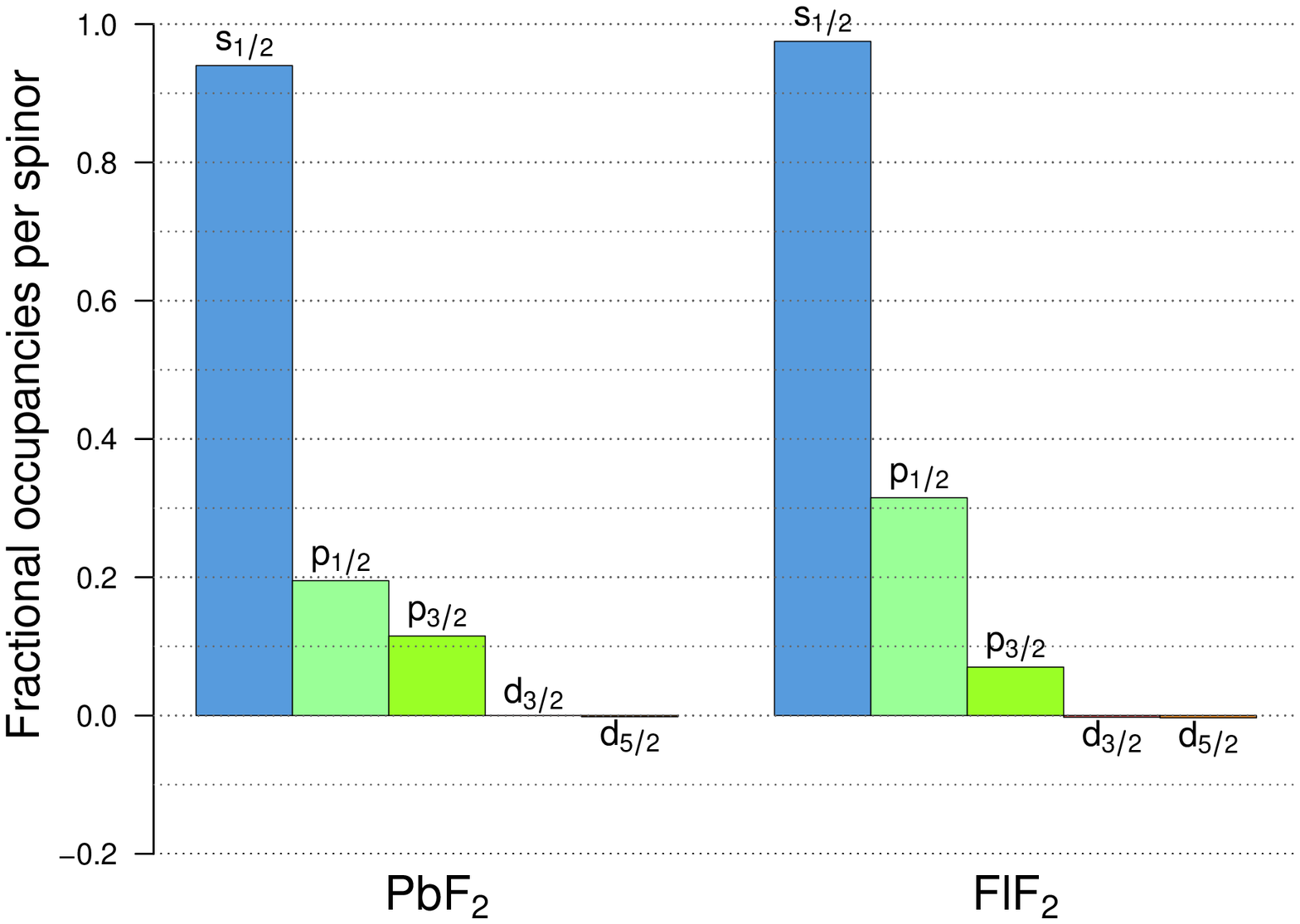}
\end{figure}

The results of the population analysis for the compounds of group 13 and 14 elements 
(see figures 1 and 2) are consistent with the simple chemical intuition: subvalence $d$- and valence $s$-shells are actually quasi inactive in bond formation, and all these elements (Tl, Nh, Pb, Fl) exhibit typical $p$-element behavior. Again, from Pb to Fl and from Tl to Nh, mixing between the subshells with the same $l$ and different $j$ becomes less efficient, and the significantly populated Fl $p_{1/2}$ subshell in FlO and $\rm FlF_2$ cannot greatly contribute to covalent bonding due to the imbalance of $p_{1/2}$ and $p_{3/2}$ occupancies. 
This leads to a weakening of bonds involving nihonium and flerovium atoms with respect to those formed by Tl and Pb, respectively.
This conclusion is also consistent with the fact that calculated bond energies decrease dramatically  from Tl and Pb to their superheavy analogues (see Table~\ref{tab:general}). For example, M--F bonds in the $\rm FlF_2$ molecule are nearly two times weaker than in $\rm PbF_2$ (2.2 eV \textit{vs} 4.2 eV).

Both PA and AiC techniques clearly shows that $d$-spinors of Pb and Fl are completely inactive in bond formation even in tetrafluorides $\rm MF_4$, where these elements exhibit their maximum oxidation state +4.
Essentially the same feature is observed for Pb and Fl oxides.

\section{Conclusions}

The application of the AiC effective configurations concept to the compounds of Hg, Cn, Tl, Nh, Pb and Fl with lighter elements gives some insight into the peculiarities of effective atomic configurations of superheavy element atoms in compounds. The decrease of the covalent component of bonds in Cn compound with respect to their Hg-containing counterparts is related to the increase of the $s$-subshell population approaching that of the closed shell. Furthermore, the large difference of hole populations of $d_{3/2}$ and $d_{5/2}$ subshells prevents their admixture necessary for strong $\sigma$-bonding. A similar imbalance of $p_{1/2}$ and $p_{3/2}$ occupancies results in significant weakening of covalent bonding in Nh and Fl compounds. The separate determination of effective populations for the subshells with the same orbital angular momentum but different total angular momenta is thus crucial for interpreting the bonding patterns in superheavy element compounds.

To summarize, local analysis of molecular Kohn--Sham density matrices in the vicinities of heavy nuclei yields essentially the same pattern of relativistic effective configurations and hence chemical bonding features.
So, both techniques can be recommended as a promising tools for interpreting the results of DFT calculations on the heaviest elements compounds in ``chemical'' terms.

This study was supported by the Russian Science Foundation (grant no. 14-31-00022).


\bibliographystyle{apsrev}
\bibliography{aicbibfile}

\begin{thebibliography}{34}
\expandafter\ifx\csname natexlab\endcsname\relax\def\natexlab#1{#1}\fi
\expandafter\ifx\csname bibnamefont\endcsname\relax
  \def\bibnamefont#1{#1}\fi
\expandafter\ifx\csname bibfnamefont\endcsname\relax
  \def\bibfnamefont#1{#1}\fi
\expandafter\ifx\csname citenamefont\endcsname\relax
  \def\citenamefont#1{#1}\fi
\expandafter\ifx\csname url\endcsname\relax
  \def\url#1{\texttt{#1}}\fi
\expandafter\ifx\csname urlprefix\endcsname\relax\def\urlprefix{URL }\fi
\providecommand{\bibinfo}[2]{#2}
\providecommand{\eprint}[2][]{\url{#2}}

\bibitem[{\citenamefont{Oganessian}(2009)}]{Oganessian2009}
\bibinfo{author}{\bibfnamefont{Y.}~\bibnamefont{Oganessian}},
  \bibinfo{journal}{Eur. Phys. J. A} \textbf{\bibinfo{volume}{42}},
  \bibinfo{pages}{361} (\bibinfo{year}{2009}), ISSN \bibinfo{issn}{1434-601X}.

\bibitem[{\citenamefont{Guseva}(2005)}]{Guseva2005}
\bibinfo{author}{\bibfnamefont{L.~I.} \bibnamefont{Guseva}},
  \bibinfo{journal}{Rus. Chem. Rev.} \textbf{\bibinfo{volume}{74}},
  \bibinfo{pages}{443} (\bibinfo{year}{2005}).

\bibitem[{\citenamefont{Eichler}(2013)}]{Eichler2013}
\bibinfo{author}{\bibfnamefont{R.}~\bibnamefont{Eichler}}, \bibinfo{journal}{J.
  Phys. Conf. Ser.} \textbf{\bibinfo{volume}{420}}, \bibinfo{pages}{012003}
  (\bibinfo{year}{2013}).

\bibitem[{\citenamefont{Sch{\"a}del and Shaughnessy}(2014)}]{Superheavy}
\bibinfo{author}{\bibfnamefont{M.}~\bibnamefont{Sch{\"a}del}} \bibnamefont{and}
  \bibinfo{author}{\bibfnamefont{D.}~\bibnamefont{Shaughnessy}},
  \emph{\bibinfo{title}{The {C}hemistry of {S}uperheavy {E}lements}}
  (\bibinfo{publisher}{Springer-Verlag Berlin Heidelberg},
  \bibinfo{year}{2014}), \bibinfo{edition}{2nd} ed.

\bibitem[{\citenamefont{Mulliken}(1955)}]{Mulliken1955}
\bibinfo{author}{\bibfnamefont{R.~S.} \bibnamefont{Mulliken}},
  \bibinfo{journal}{J. Chem. Phys.} \textbf{\bibinfo{volume}{23}},
  \bibinfo{pages}{1833} (\bibinfo{year}{1955}).

\bibitem[{\citenamefont{Pershina et~al.}(2002)\citenamefont{Pershina, Bastug,
  Jacob, Fricke, and Varga}}]{Pershina2002}
\bibinfo{author}{\bibfnamefont{V.}~\bibnamefont{Pershina}},
  \bibinfo{author}{\bibfnamefont{T.}~\bibnamefont{Bastug}},
  \bibinfo{author}{\bibfnamefont{T.}~\bibnamefont{Jacob}},
  \bibinfo{author}{\bibfnamefont{B.}~\bibnamefont{Fricke}}, \bibnamefont{and}
  \bibinfo{author}{\bibfnamefont{S.}~\bibnamefont{Varga}},
  \bibinfo{journal}{Chem. Phys. Lett.} \textbf{\bibinfo{volume}{365}},
  \bibinfo{pages}{176 } (\bibinfo{year}{2002}), ISSN \bibinfo{issn}{0009-2614}.

\bibitem[{\citenamefont{Jensen}(2007)}]{Jensen}
\bibinfo{author}{\bibfnamefont{F.}~\bibnamefont{Jensen}},
  \emph{\bibinfo{title}{Introduction to {C}omputational {C}hemistry}}
  (\bibinfo{publisher}{John Wiley \& Sons}, \bibinfo{year}{2007}),
  \bibinfo{edition}{3rd} ed.

\bibitem[{\citenamefont{Dubillard et~al.}(2006)\citenamefont{Dubillard, Rota,
  Saue, and Faegri}}]{Dubillard2006}
\bibinfo{author}{\bibfnamefont{S.}~\bibnamefont{Dubillard}},
  \bibinfo{author}{\bibfnamefont{J.-B.} \bibnamefont{Rota}},
  \bibinfo{author}{\bibfnamefont{T.}~\bibnamefont{Saue}}, \bibnamefont{and}
  \bibinfo{author}{\bibfnamefont{K.}~\bibnamefont{Faegri}},
  \bibinfo{journal}{J. Chem. Phys.} \textbf{\bibinfo{volume}{124}},
  \bibinfo{pages}{154307} (\bibinfo{year}{2006}).

\bibitem[{\citenamefont{Titov et~al.}(2014)\citenamefont{Titov, Lomachuk, and
  Skripnikov}}]{TitovLomachuk2014}
\bibinfo{author}{\bibfnamefont{A.~V.} \bibnamefont{Titov}},
  \bibinfo{author}{\bibfnamefont{Y.~V.} \bibnamefont{Lomachuk}},
  \bibnamefont{and} \bibinfo{author}{\bibfnamefont{L.~V.}
  \bibnamefont{Skripnikov}}, \bibinfo{journal}{Phys. Rev. A}
  \textbf{\bibinfo{volume}{90}}, \bibinfo{pages}{052522}
  (\bibinfo{year}{2014}).

\bibitem[{\citenamefont{Zaitsevskii et~al.}(2016)\citenamefont{Zaitsevskii,
  Skripnikov, and Titov}}]{Zaitsevskii2016}
\bibinfo{author}{\bibfnamefont{A.~V.} \bibnamefont{Zaitsevskii}},
  \bibinfo{author}{\bibfnamefont{L.~V.} \bibnamefont{Skripnikov}},
  \bibnamefont{and} \bibinfo{author}{\bibfnamefont{A.~V.} \bibnamefont{Titov}},
  \bibinfo{journal}{Mendeleev Commun.} \textbf{\bibinfo{volume}{26}},
  \bibinfo{pages}{307 } (\bibinfo{year}{2016}), ISSN \bibinfo{issn}{0959-9436}.

\bibitem[{\citenamefont{Pershina et~al.}(2009)\citenamefont{Pershina, Anton,
  and Jacob}}]{Pershina2009}
\bibinfo{author}{\bibfnamefont{V.}~\bibnamefont{Pershina}},
  \bibinfo{author}{\bibfnamefont{J.}~\bibnamefont{Anton}}, \bibnamefont{and}
  \bibinfo{author}{\bibfnamefont{T.}~\bibnamefont{Jacob}}, \bibinfo{journal}{J.
  Chem. Phys.} \textbf{\bibinfo{volume}{131}}, \bibinfo{pages}{084713}
  (\bibinfo{year}{2009}).

\bibitem[{\citenamefont{Dmitriev et~al.}(2014)\citenamefont{Dmitriev, Aksenov,
  Albin, Bozhikov, Chelnokov, Chepygin, Eichler, Isaev, Katrasev, Lebedev
  et~al.}}]{Dmitriev2014}
\bibinfo{author}{\bibfnamefont{S.~N.} \bibnamefont{Dmitriev}},
  \bibinfo{author}{\bibfnamefont{N.~V.} \bibnamefont{Aksenov}},
  \bibinfo{author}{\bibfnamefont{Y.~V.} \bibnamefont{Albin}},
  \bibinfo{author}{\bibfnamefont{G.~A.} \bibnamefont{Bozhikov}},
  \bibinfo{author}{\bibfnamefont{M.~L.} \bibnamefont{Chelnokov}},
  \bibinfo{author}{\bibfnamefont{V.~I.} \bibnamefont{Chepygin}},
  \bibinfo{author}{\bibfnamefont{R.}~\bibnamefont{Eichler}},
  \bibinfo{author}{\bibfnamefont{A.~V.} \bibnamefont{Isaev}},
  \bibinfo{author}{\bibfnamefont{D.~E.} \bibnamefont{Katrasev}},
  \bibinfo{author}{\bibfnamefont{V.~Y.} \bibnamefont{Lebedev}},
  \bibnamefont{et~al.}, \bibinfo{journal}{Mendeleev Commun.}
  \textbf{\bibinfo{volume}{24}}, \bibinfo{pages}{253 } (\bibinfo{year}{2014}),
  ISSN \bibinfo{issn}{0959-9436}.

\bibitem[{\citenamefont{Oleynichenko and Zaitsevskii}(2017)}]{Oleynichenko2017}
\bibinfo{author}{\bibfnamefont{A.}~\bibnamefont{Oleynichenko}}
  \bibnamefont{and}
  \bibinfo{author}{\bibfnamefont{A.}~\bibnamefont{Zaitsevskii}},
  \bibinfo{journal}{Nonlinear Phenomena in Complex Systems}
  \textbf{\bibinfo{volume}{20}}, \bibinfo{pages}{177 } (\bibinfo{year}{2017}).

\bibitem[{\citenamefont{Knizia}(2013)}]{Knizia2013}
\bibinfo{author}{\bibfnamefont{G.}~\bibnamefont{Knizia}}, \bibinfo{journal}{J.
  Chem. Theory Comput.} \textbf{\bibinfo{volume}{9}}, \bibinfo{pages}{4834}
  (\bibinfo{year}{2013}).

\bibitem[{\citenamefont{Titov and Mosyagin}(1999)}]{Proportion}
\bibinfo{author}{\bibfnamefont{A.}~\bibnamefont{Titov}} \bibnamefont{and}
  \bibinfo{author}{\bibfnamefont{N.}~\bibnamefont{Mosyagin}},
  \bibinfo{journal}{Int. J. Quantum Chem.} \textbf{\bibinfo{volume}{71}},
  \bibinfo{pages}{359} (\bibinfo{year}{1999}).

\bibitem[{\citenamefont{Adamo and Barone}(1999)}]{Adamo1999}
\bibinfo{author}{\bibfnamefont{C.}~\bibnamefont{Adamo}} \bibnamefont{and}
  \bibinfo{author}{\bibfnamefont{V.}~\bibnamefont{Barone}},
  \bibinfo{journal}{J. Chem. Phys.} \textbf{\bibinfo{volume}{110}},
  \bibinfo{pages}{6158} (\bibinfo{year}{1999}).

\bibitem[{\citenamefont{Demidov and Zaitsevskii}(2015)}]{Demidov2015}
\bibinfo{author}{\bibfnamefont{Y.}~\bibnamefont{Demidov}} \bibnamefont{and}
  \bibinfo{author}{\bibfnamefont{A.}~\bibnamefont{Zaitsevskii}},
  \bibinfo{journal}{Chem. Phys. Lett.} \textbf{\bibinfo{volume}{638}},
  \bibinfo{pages}{21 } (\bibinfo{year}{2015}).

\bibitem[{\citenamefont{Mosyagin et~al.}(2010)\citenamefont{Mosyagin,
  Zaitsevskii, and Titov}}]{Mosyagin2010}
\bibinfo{author}{\bibfnamefont{N.~S.} \bibnamefont{Mosyagin}},
  \bibinfo{author}{\bibfnamefont{A.}~\bibnamefont{Zaitsevskii}},
  \bibnamefont{and} \bibinfo{author}{\bibfnamefont{A.~V.} \bibnamefont{Titov}},
  \bibinfo{journal}{Int. Rev. At. Mol. Phys.} \textbf{\bibinfo{volume}{1}},
  \bibinfo{pages}{63} (\bibinfo{year}{2010}), ISSN \bibinfo{issn}{2229-3159}.

\bibitem[{pnp()}]{pnpi_site}
\emph{\bibinfo{title}{Effective potentials and basis sets}},
  \urlprefix\url{http://www.qchem.pnpi.spb.ru/recp.html}.

\bibitem[{\citenamefont{Rappoport and Furche}(2010)}]{Rappoport2010}
\bibinfo{author}{\bibfnamefont{D.}~\bibnamefont{Rappoport}} \bibnamefont{and}
  \bibinfo{author}{\bibfnamefont{F.}~\bibnamefont{Furche}},
  \bibinfo{journal}{J. Chem. Phys.} \textbf{\bibinfo{volume}{133}},
  \bibinfo{pages}{134105} (\bibinfo{year}{2010}).

\bibitem[{DIR()}]{DIRAC15}
\bibinfo{note}{DIRAC, a relativistic ab initio electronic structure program,
  Release DIRAC15 (2015), written by R. Bast, T. Saue, L. Visscher, and H. J.
  Aa. Jensen, with contributions from V. Bakken, K. G. Dyall, S. Dubillard, U.
  Ekstroem, E. Eliav, T. Enevoldsen, E. Fasshauer, T. Fleig, O. Fossgaard, A.
  S. P. Gomes, T. Helgaker, J. Henriksson, M. Ilias, Ch. R. Jacob, S. Knecht,
  S. Komorovsky, O. Kullie, J. K. Laerdahl, C. V. Larsen, Y. S. Lee, H. S.
  Nataraj, M. K. Nayak, P. Norman, G. Olejniczak, J. Olsen, Y. C. Park, J. K.
  Pedersen, M. Pernpointner, R. Di Remigio, K. Ruud, P. Salek, B.
  Schimmelpfennig, J. Sikkema, A. J. Thorvaldsen, J. Thyssen, J. van Stralen,
  S. Villaume, O. Visser, T. Winther, and S. Yamamoto (see
  http://www.diracprogram.org).}

\bibitem[{\citenamefont{van W\"{u}llen}(2010)}]{Wullen2010}
\bibinfo{author}{\bibfnamefont{C.}~\bibnamefont{van W\"{u}llen}},
  \bibinfo{journal}{Z. Phys. Chem.} \textbf{\bibinfo{volume}{224}},
  \bibinfo{pages}{413} (\bibinfo{year}{2010}).

\bibitem[{\citenamefont{Tang et~al.}(2009)\citenamefont{Tang, Sanville, and
  Henkelman}}]{Henkelman2009}
\bibinfo{author}{\bibfnamefont{W.}~\bibnamefont{Tang}},
  \bibinfo{author}{\bibfnamefont{E.}~\bibnamefont{Sanville}}, \bibnamefont{and}
  \bibinfo{author}{\bibfnamefont{G.}~\bibnamefont{Henkelman}},
  \bibinfo{journal}{J. Phys. Condens. Matter} \textbf{\bibinfo{volume}{21}},
  \bibinfo{pages}{084204} (\bibinfo{year}{2009}).

\bibitem[{\citenamefont{Sanville et~al.}(2007)\citenamefont{Sanville, Kenny,
  Smith, and Henkelman}}]{SanvilleHenkelman2007}
\bibinfo{author}{\bibfnamefont{E.}~\bibnamefont{Sanville}},
  \bibinfo{author}{\bibfnamefont{S.~D.} \bibnamefont{Kenny}},
  \bibinfo{author}{\bibfnamefont{R.}~\bibnamefont{Smith}}, \bibnamefont{and}
  \bibinfo{author}{\bibfnamefont{G.}~\bibnamefont{Henkelman}},
  \bibinfo{journal}{J. Comp. Chem.} \textbf{\bibinfo{volume}{28}},
  \bibinfo{pages}{899} (\bibinfo{year}{2007}).

\bibitem[{\citenamefont{Skripnikov and Titov}(2015)}]{Skripnikov2015}
\bibinfo{author}{\bibfnamefont{L.}~\bibnamefont{Skripnikov}} \bibnamefont{and}
  \bibinfo{author}{\bibfnamefont{A.}~\bibnamefont{Titov}},
  \bibinfo{journal}{Phys. Rev. A} \textbf{\bibinfo{volume}{91}},
  \bibinfo{pages}{042504} (\bibinfo{year}{2015}).

\bibitem[{\citenamefont{Shepler and Peterson}(2003)}]{SheplerPeterson2003}
\bibinfo{author}{\bibfnamefont{B.~C.} \bibnamefont{Shepler}} \bibnamefont{and}
  \bibinfo{author}{\bibfnamefont{K.~A.} \bibnamefont{Peterson}},
  \bibinfo{journal}{J. Phys. Chem. A} \textbf{\bibinfo{volume}{107}},
  \bibinfo{pages}{1783 } (\bibinfo{year}{2003}).

\bibitem[{\citenamefont{Liu et~al.}(1999)\citenamefont{Liu, Franke, and
  Dolg}}]{Liu1999}
\bibinfo{author}{\bibfnamefont{W.}~\bibnamefont{Liu}},
  \bibinfo{author}{\bibfnamefont{R.}~\bibnamefont{Franke}}, \bibnamefont{and}
  \bibinfo{author}{\bibfnamefont{M.}~\bibnamefont{Dolg}},
  \bibinfo{journal}{Chem. Phys. Lett.} \textbf{\bibinfo{volume}{302}},
  \bibinfo{pages}{231 } (\bibinfo{year}{1999}), ISSN \bibinfo{issn}{0009-2614}.

\bibitem[{\citenamefont{Seth et~al.}(1997)\citenamefont{Seth, Schwerdtfeger,
  and Dolg}}]{Seth1997}
\bibinfo{author}{\bibfnamefont{M.}~\bibnamefont{Seth}},
  \bibinfo{author}{\bibfnamefont{P.}~\bibnamefont{Schwerdtfeger}},
  \bibnamefont{and} \bibinfo{author}{\bibfnamefont{M.}~\bibnamefont{Dolg}},
  \bibinfo{journal}{J. Chem. Phys.} \textbf{\bibinfo{volume}{106}},
  \bibinfo{pages}{3623 } (\bibinfo{year}{1997}).

\bibitem[{\citenamefont{Lide}(2000)}]{CRC}
\bibinfo{editor}{\bibfnamefont{D.~P.} \bibnamefont{Lide}}, ed.,
  \emph{\bibinfo{title}{{CRC Handbook of Chemistry and Physics}}}
  (\bibinfo{publisher}{CRC PRESS}, \bibinfo{year}{2000}),
  \bibinfo{edition}{81st} ed.

\bibitem[{\citenamefont{Liu et~al.}(2001)\citenamefont{Liu, van Wuellen, Han,
  Choi, and Lee}}]{Liu2001}
\bibinfo{author}{\bibfnamefont{W.}~\bibnamefont{Liu}},
  \bibinfo{author}{\bibfnamefont{C.}~\bibnamefont{van Wuellen}},
  \bibinfo{author}{\bibfnamefont{Y.~K.} \bibnamefont{Han}},
  \bibinfo{author}{\bibfnamefont{Y.~J.} \bibnamefont{Choi}}, \bibnamefont{and}
  \bibinfo{author}{\bibfnamefont{Y.~S.} \bibnamefont{Lee}}, in
  \emph{\bibinfo{booktitle}{New Perspectives in Quantum Systems in Chemistry
  and Physics, Part 1}} (\bibinfo{publisher}{Academic Press},
  \bibinfo{year}{2001}), vol.~\bibinfo{volume}{39} of
  \emph{\bibinfo{series}{Advances in Quantum Chemistry}}, pp.
  \bibinfo{pages}{325 -- 355}.

\bibitem[{\citenamefont{{Chase Jr.}}(1998)}]{ChaseTables1998}
\bibinfo{author}{\bibfnamefont{M.~W.} \bibnamefont{{Chase Jr.}}},
  \bibinfo{journal}{J. Phys. Chem. Ref. Data, Monograph 9} pp.
  \bibinfo{pages}{1 -- 1951} (\bibinfo{year}{1998}).

\bibitem[{\citenamefont{Seth et~al.}(1998)\citenamefont{Seth, Faegri, and
  Schwerdtfeger}}]{Seth1998}
\bibinfo{author}{\bibfnamefont{M.}~\bibnamefont{Seth}},
  \bibinfo{author}{\bibfnamefont{K.}~\bibnamefont{Faegri}}, \bibnamefont{and}
  \bibinfo{author}{\bibfnamefont{P.}~\bibnamefont{Schwerdtfeger}},
  \bibinfo{journal}{Angew. Chem. Int. Ed.} \textbf{\bibinfo{volume}{37}},
  \bibinfo{pages}{2493 } (\bibinfo{year}{1998}).

\bibitem[{PbF(2006)}]{PbF4expt}
in \emph{\bibinfo{booktitle}{Encyclopedia of Inorganic Chemistry}}
  (\bibinfo{publisher}{John Wiley \& Sons, Ltd}, \bibinfo{year}{2006}), ISBN
  \bibinfo{isbn}{9780470862100}.

\bibitem[{\citenamefont{Lee}(2004)}]{Lee2004}
\bibinfo{author}{\bibfnamefont{Y.~S.} \bibnamefont{Lee}}, in
  \emph{\bibinfo{booktitle}{Relativistic Electronic Structure Theory}}, edited
  by \bibinfo{editor}{\bibfnamefont{P.}~\bibnamefont{Schwerdtfeger}}
  (\bibinfo{publisher}{Elsevier}, \bibinfo{year}{2004}),
  vol.~\bibinfo{volume}{14} of \emph{\bibinfo{series}{Theoretical and
  Computational Chemistry}}, pp. \bibinfo{pages}{352 -- 416}.

\end{thebibliography}

\end{document}